\documentclass[amssymb,aps,preprint]{revtex4}
\usepackage{graphicx}
\usepackage{latexsym}
\usepackage[dvips]{color}
\usepackage[normalem]{ulem}

\def\refpar[#1]{(\ref{#1})}
\def\msd{{\rm m.s.d.}~}
\newcommand{\beq}{\begin{equation}}
\newcommand{\eeq}{\end{equation}}
\newcommand{\beqa}{\begin{eqnarray}}
\newcommand{\eeqa}{\end{eqnarray}}
\newcommand{\ba}{\begin{array}}
\newcommand{\ea}{\end{array}}

\begin{document}

\title{Single file diffusion of macroscopic charged particles}

 \author{C. Coste, J.-B. Delfau, C. Even and M. Saint Jean}
 \affiliation{ Laboratoire MSC. UMR CNRS 7057 et Universit\'e Paris Diderot -Paris7  \\B\^atiment Condorcet 
  10 rue Alice Domon et L\'eonie Duquet  \\  75205  PARIS Cedex 13 -  France
}
 \date{\today}

\begin{abstract}

In this paper, we study a macroscopic system of electrically interacting metallic beads organized as a sequence along an annulus. A random mechanical shaking mimics the thermal excitation. We exhibit non Fickian diffusion (Single File Diffusion) at large time. We measure the mobility of the particles, and compare it to theoretical expectations. We show that our system cannot be accurately described by theories assuming only hard sphere interactions. Its behavior is qualitatively described by a theory extended to more realistic potentials [Kollmann, PRL {\bf 90} 180602, (2003)]. A correct quantitative agreement is shown, and we interpret the discrepancies by the violation of a key assumption of the theory, that of overdamped dynamics. We recast previous results on colloids with known interaction potentials, and compare them quantitatively to the theory. Focusing on the transition between ordinary and single file diffusion, we exhibit a dimensionless crossover time that is of order one both for colloids and our system, although the time and length scales differ by several orders of magnitude.

PACS~: 05.40.-a	Fluctuation phenomena, random processes, noise, and Brownian motion
66.10.cg	Mass diffusion, including self-diffusion, mutual diffusion, tracer diffusion, etc. 
47.57.eb	Diffusion and aggregation

\end{abstract}

\maketitle

\section{Introduction}

When particles transport occurs in highly confined geometries, in such  a way that they are distributed along a line, and contained in channels so narrow that they cannot cross each others, very strong correlations appear between the particles, and diffusive transport is anomalous. The particles are said to undergo Single File Diffusion (SFD). Such occurences include transport of water and ions through molecular-sized channels in biological membranes \cite{Hodkin55,Rosenberg78,Hernandez92}, molecular sieving effects in nanoporous zeolites \cite{Gupta95, Hahn96, Sholl97}, diffusion of molecules along folded polymer chains \cite{Wang09} and diffusion of colloids in microfluidic devices \cite{Wei00,Lin05} or in specially designed channels generated by scanning optical tweezers \cite{LutzJPC04,LutzPRL04}.

SFD is a long standing problem in probability theory. The simplest modelisation consider the particles as hard spheres, interacting with a zero range singular potential \cite{Jepsen65, Arratia83, deGennes71, Levitt73, Richards77, vanBeijeren83, Karger92, Hahn95, Hahn98,Taloni06}, and revelled the prominent effect of the confinement~: The diffusion of a given particle is subdiffusive, with the mean-squared displacement $\langle \Delta x^2 \rangle$ that scales at large times $t$ with a non-fickian exponent,
\begin{equation}
\langle \Delta x^2 \rangle = F  {t}^{1/2},
\label{eq:SFD}
\end{equation}
$F$ beeing called the mobility. For hard core interactions between particles, the mobility $F_H$ reads
\begin{equation}
F_H = {2\over \rho}\sqrt{D_0 \over \pi},
\label{eq:mobilhard}
\end{equation}
where $\rho$ is the number of particles per unit length and $D_0$ the single particle diffusivity.

This non-fickian scaling at large time has already been observed for molecular transport in zeolites \cite{Gupta95,Hahn96,Sholl97}, colloids ine one-dimensional channels \cite{Wei00,LutzJPC04,LutzPRL04,Lin05}, charged macroscopic beads in circular channels \cite{Coupier06,Coupier07} and in numerical simulations of brownian particles interacting with a finite range pair potential \cite{Herrera07,Nelissen07,Herrera08}. When the focus is put on the mobility itself, some of those works show discrepancies with the hard sphere description  \refpar[eq:mobilhard]. Gupta \emph{et al.} measured the mobility of ethane in $AlPO_4-5$, and found it one order of magnitude greater than expected \cite{Gupta95}. For colloids, the measurements show a dependence of the mobility on the strength of the interparticle potential \cite{Wei00,Herrera07,Herrera08}, in contrast with \refpar[eq:mobilhard]. 

Recently, Kollmann \cite{Kollmann03} extended SFD theory to brownian particles interacting with a finite range pair potential. His work relies on three fundamental assumptions. First, the calculations are valid in the thermodynamic limit, for infinite systems. Then, at the very basis of the calculation, it is assumed that the dynamics of the brownian particles is overdamped. Lastly, the hydrodynamic interactions, which are the feedback effect on a given particle motion that comes from the flow of the surrounding fluid induced by that very motion, are neglected. In this paper, we want to address the relevance of those assumptions to actual instances of SFD. We discuss below finite size effects and overdamping, and we review previous works in which hydrodynamic interactions could be an issue  \cite{Wei00}. 

Another issue is the dependence of SFD mobility on the temperature, brownian particles density and interaction strength. The evolution of the mobility with the density, for finite range interactions between the particles, have been checked for colloids by Lutz \emph{et al.} \cite{LutzJPC04,LutzPRL04}, and by Lin \emph{et al.}  \cite{Lin05}, who found good agreement with Kollmann's calculation. It was pointed out in \cite{Lin05} that some colloids behave as a Tonks gas \cite{Tonks36}, \emph{i.e.} a gas of hard rods, for which Kollmann's result is equivalent to the classical hard sphere model, rescaling the density to take into account the size of the particles. Other authors have varied the strength of the interactions between the brownian particles  in experiments  \cite{Wei00} and simulations \cite{Herrera07,Herrera08}. They observed the dependence of the mobility on the interaction strength but do not compare their results to Kollmann's calculation. We review their works, recasting their data to allow quantitative comparison with theoretical predictions. Then, we discuss our own experimental observations of SFD in a system of macroscopically charged particles in which the temperature, particle density and interaction strength may be varied independently. The interaction potential in our system is well described by a modified Bessel function $K_0$ [see \cite{Galatola06} and eqn.~\refpar[eq:gammabeads] below], which is of the same functional shape as vortices interactions in superconductors. This study may thus be relevant to vortex transport in mesoscopic channels in superconductors \cite{Besseling99,Anders00,Kokubo04}.

The paper is organized in the following way. In section~\ref{sec:soa}, we review the theoretical predictions for SFD with finite range potentials between the diffusing particles (\S~\ref{sec:theory}), describe the relevant interaction potentials and recall their link with the isothermal compressibility (\S~\ref{sec:potentials}), and compare the observations to the predictions of Kollmann \cite{Kollmann03} in \ref{sec:previous}. Then, in section~\ref{sec:process}, we describe our experimental system (\S~\ref{sec:setup}), insisting on the random mechanical shaking applied to the system, that behaves as an effective thermodynamic temperature (\S~\ref{sec:temp}). In section~\ref{sec:results}, we set out our results, providing evidences of SFD for macroscopic charged beads (\S~\ref{sec:sfd}), and studying the evolution of the mobility with the strength of interparticle potential, particle density and (effective) thermodynamic temperature (\S~\ref{sec:mobil}). We discuss potential finite size effects (\S~\ref{sec:size}) and put peculiar emphasis on the crossover time between normal diffusion and SFD (\S~\ref{sec:crossover}). A last section~\ref{sec:conclusion} is devoted to a summary of our results and our conclusions.

\section{Single file diffusion~: A state of the art.}
\label{sec:soa}

\subsection{Single file diffusion for finite range interparticle potential}
\label{sec:theory}

Recently, Kollmann \cite{Kollmann03} have extended SFD to more realistic, finite range interactions between the diffusing particles. The subdiffusive behavior, with a mean-squared displacement (hereafter "\msd")  scaling at large times $t$ with a non Fickian exponent $1/2$ as in \refpar[eq:mobilhard], is recovered. However, the mobility is now given by
\begin{equation}
F_S= {2\over \rho} S(0,0)\sqrt{{ D_{eff} \over \pi}},
\label{eq:mobilsoftdeb}
\end{equation}
where we have introduced the static structure factor of the particles, in the limit of long wavelength or zero wavenumber $q$, $S(0,0) \equiv S(q \to 0,t = 0)$. The diffusivity $D_{eff}$ is the  effective diffusivity of a brownian particle, taking into account its interactions with the other brownian particles, and differs from the single particle diffusivity $D_0$. Independently from peculiar effects linked to SFD geometry, interparticle interactions lead to a renormalization of the single particle diffusivity  \cite{Nagele96}. When hydrodynamic interactions are neglected, the effective diffusivity $D_{eff} = D_0/S(0,0)$ \cite{Nagele96}. With this expression, the mobility of brownian particles with finite range interactions reads
\begin{equation}
F_S= {2\over \rho} \sqrt{{S(0,0) D_{0} \over \pi}},
\label{eq:mobilsoft}
\end{equation}

The static structure factor may be measured by light diffusion for suspensions, or calculated from the particles positions in simulations. It may also be expressed in terms of the isothermal compressibility $\kappa_T$ as $S(0,0) = \rho k_B T \kappa_T$, where $k_B$ is the Boltzmann constant and $T$ the thermodynamic temperature  \cite{Nagele96}. This is particularly well suited to the analysis of our experiments, since we have rather small systems and a poor resolution on structure factor measurements. In contrast, the interaction potential is known with a good precision, and allows an easy determination of $\kappa_T$. 

Generically, the interaction potential may be expressed as $U(r) = U_0 f(r/\lambda)$ where $\lambda$ is some characteristic length, $U_0$ some energy scale and $r$ the distance between the particles.  For interacting particles on a line with a mean interparticle distance $\langle r \rangle = 1/\rho$, we may calculate the isothermal compressibility at zero temperature \cite{Brillouin53}, 
\begin{equation}
\kappa_T ={\rho \over \sum\limits_{m>0}m^2 U'' \left({m\over \rho}\right)} = {\rho \lambda^2 \over U_0 \sum\limits_{m>0}m^2 f'' \left({m\over \lambda\rho}\right)} ,
\label{eq:compressibility}
\end{equation}
where the summation extends, in principle, to all particules. In practice, for the potentials that we consider, the sum converges fastly and keeping only three terms in the sum already gives four significant figures. The mobility thus reads
\beq
F_S  = {2  \over {\rho}}\sqrt{{D_0 {\rho} k_B T \kappa_T \over \pi}} = {2}\left[{D_0 k_B T \lambda^2 \over \pi U_0 \sum\limits_{m>0} m^2f''\left({m\over \lambda\overline{\rho} }\right)}\right]^{1/2}.
\label{eq:mobilsoftfin}
\eeq
This result is valid for a suspension of brownian particles interacting with a pair potential that is a function of interparticles distance only, when hydrodynamic interactions due to the coherent solvent motions  are neglected. In principle, this calculation does not include entropic contributions since it is valid at zero temperature, but finite temperature corrections are small \cite{Kleinert89}.

For the sake of comparison with other works \cite{Wei00,Herrera07,Herrera08}, we recall the definition of the reduced mobility $F^* \equiv \rho F/\sqrt{D_0}$, which for hard sphere potential is a constant equal to $2/\sqrt{\pi}$~\refpar[eq:SFD]. We also introduce the dimensionless potential $\Gamma \equiv \beta U_0 f(1/\lambda \rho)$ where $\beta = 1/( k_B T)$. With those variables, the reduced mobility may be writen
\begin{equation}
F_S^*= {2\over\sqrt{ \pi \Gamma}} {\rho\lambda \over \left[ \sum\limits_{m>0}m^2 f''\left({m\over \lambda\rho}\right)/f\left({1\over \lambda\rho}\right)\right]^{1/2}}.
\label{eq:mobreduced}
\end{equation}
It is clear from this expression that, although dimensionless, the variables $(\Gamma,F^*)$ are not scaling variables since there remains a dependence of the reduced mobility on the density $\rho$.

\subsection{Interaction potentials}
\label{sec:potentials}

For the sake of convenience, we recall here the interaction potentials of concern in what follows~:

(a) the interaction between magnetic moments induced in a brownian suspension of paramagnetic particles by an applied magnetic field $B$  \cite{Wei00,Herrera08}. The dimensionless potential energy $\Gamma_{\rm para}$ reads
\begin{equation}
\Gamma_{\rm para} \equiv \beta {\mu_0 \over 4 \pi}{\chi_{eff}^2 B^2 \over \lambda^3}\left({\lambda \over r}\right)^3,
\label{eq:gammapara}
\end{equation}
with $\mu_0$ is the vacuum permeability, $\chi_{eff}$ the magnetic suceptibility of the particles, and where the characteristic length $\lambda$ may be, \emph{e.g.} the particle diameter $\sigma$. The susceptibility that corresponds to the experiments \cite{Wei00} is $2.2\times 10^{-12}$A m$^2$T$^{-1}$, the same value beeing used in the simulations \cite{Herrera08}.

(b) a screened electrostatic interaction described by a DLVO pair potential (see \cite{Nagele96}, appendix A), with dimensionless potential energy $\Gamma_{\rm DLVO} $
\begin{equation}
\Gamma_{\rm DLVO} \equiv Z_{eff}^2{\lambda_B\over \lambda}\left({e^{\sigma/(2  \lambda)}\over  1 + \sigma/(2  \lambda)}\right)^2 {e^{- r/ \lambda} \over r/ \lambda},
\label{eq:gammaDLVO}
\end{equation}
where $Z_{eff}$ is the effective charge of the colloid, $\lambda_B$ the Bjerrum length, $ \lambda$ the Debye screening length and $\sigma$ the particles diameter. The relevant values of those parameters, used in the simulations \cite{Herrera07} and taken from experiments \cite{Bleil06}, are $\sigma = 2.8\;\mu$m, $Z_{eff} = 5400$, $\lambda_B = 0.72$~nm and $ \lambda = 550$~nm.

(c) the interaction energy for our experimental system of metallic beads contained in a condenser of thickness $h$ under an applied voltage $V_0$. The set-up is described below in sec.~\ref{sec:setup}.  The dimensionless pair potential $\Gamma_{\rm beads} $ reads \cite{Galatola06}, 
\begin{equation}
\Gamma_{\rm beads} =  \beta e_0 \epsilon_0 V_0^2 h K_0\left({r \over \lambda}\right), 
\label{eq:gammabeads}
\end{equation}
where $e_0 \approx 0.71$ is a numerical constant, $\epsilon_0$ the vacuum permittivity,  $K_0$ the modified Bessel function of second order. In our experiments $h = 1.5$~mm  and the characteristic length $\lambda \approx 0.32 h \approx 0.5$~mm. The parameters $e_0$ and $\lambda$ depend on the dimensions of the beads, and the geometry of the condenser but are independent of the applied voltage $V_0$ (see below sec.~\ref{sec:setup}). They are kept fixed in all our experiments. The tension $V_0$ may vary in the range $[800,1400]~V$, allowing us to easily monitor the interaction energy between the metallic beads. This interaction is similar to the screened electrical interaction between charged particles in suspensions \refpar[eq:gammaDLVO], because it depends on $r$ as $\exp(-r/\lambda)/\sqrt{r}$ for large $r$. 

\subsection{Experiments and simulations on SFD with soft interactions}
\label{sec:previous}

In this section, we discuss previous works on colloids in SFD geometry, and compare quantitatively their results to the predictions of Kollmann \cite{Kollmann03}. We have collected the existing data and plotted the measured values of the reduced mobility $F^*$ [see \refpar[eq:mobreduced]], either for varying potential strength \cite{Wei00,Herrera08} or varying packing fraction \cite{Herrera07}. In all cases, the interaction potentials are explicitely known (see \S~\ref{sec:potentials}) and equation \refpar[eq:mobreduced] allows the calculation of the reduced mobility predicted by Kollmann without any free parameter. We add to the existing mobility data the corresponding theoretical predictions, thus providing a quantitative test of the theory. We recall in Table~\ref{tab:listhypo} the relevant experimental or numerical characteristics of the systems that are discussed below and in section~\ref{sec:results}.

\begin{table}[htb]
\begin{center}
\begin{tabular}{|c|c|c|c|c|c|}
\hline
System & Reference & $U(r)$ & $N$ & Dynamics &  H.I. \\
\hline
\hline
\textbf{Theory} &\cite{Kollmann03} & \textbf{finite range} & \textbf{$\infty$} & \textbf{overdamped} & \textbf{no} \\
\hline
Colloids (num) & \cite{Herrera07} &  \refpar[eq:gammapara]  & 900 & overdamped & no \\
\hline
Colloids (exp) &\cite{Wei00} & \refpar[eq:gammapara] & 89 & overdamped & yes \\
\hline
Colloids (num) & \cite{Herrera08} & \refpar[eq:gammaDLVO]  & 900 & overdamped & no  \\
\hline
Colloids (exp) & \cite{LutzPRL04,LutzJPC04} & \refpar[eq:gammaDLVO] & 49 & overdamped & no \\
\hline
Colloids (num) & \cite{Nelissen07} &  \refpar[eq:gammaDLVO] & $\in [50,100]$ & underdamped & no \\
\hline
Charged beads (exp) & this work & \refpar[eq:gammabeads] & $\in [12,37]$ & underdamped & no \\
\hline
\end{tabular}
\caption{Characteristics of experimental (exp) and numerical (num) systems undergoing SFD. Colloid dynamics is presumably overdamped in experiments \cite{Nagele96}. H.I. stands for "Hydrodynamic Interactions".}
\label{tab:listhypo}
\end{center}
\end{table}

\subsubsection{Colloids with paramagnetic interactions}
\label{sec:para}

Wei \emph{et al.} realized single file geometries by putting colloids in circular channels etched by photolithography. The colloidal particles are polystyrene beads doped with Fe$_2$O$_3$ clusters, so that when submitted to an external magnetic field $B$ they interact  as magnetic dipoles whose moments depend on $B$, with an interaction potential given by \refpar[eq:gammapara]. It is thus possible to study the behavior of the system with varying interaction strength. The same system is simulated by Herrera-Velarde and Casta$\tilde{\rm n}$eda-Priego \cite{Herrera08}.

\begin{figure}[htb] 
\vskip-0.25cm
\begin{center}
\includegraphics{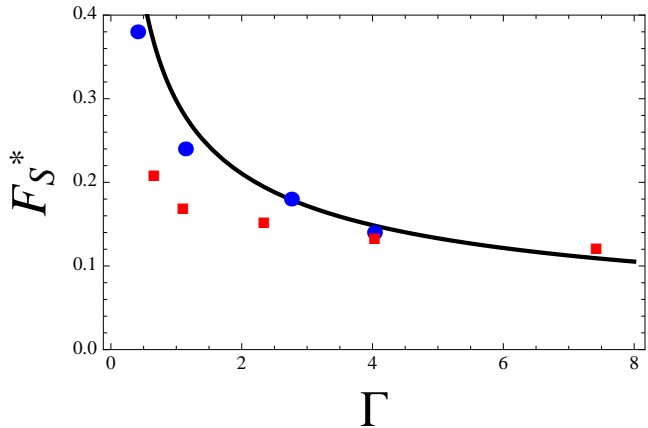}
\caption{(color online) Plot of the reduced mobility $F_S^*$ as a function of the reduced interaction $\Gamma$ for colloids with magnetic dipole interactions. \textcolor{blue}{$\bullet$}~: Numerical simulations \cite{Herrera08}, \textcolor{red}{$\blacksquare$}~: experimental data \cite{Wei00}, solid line Kollmann's formula, Eqn.~(\ref{eq:mobreduced}) with (\ref{eq:gammapara}) (no free parameter in this calculation).}
\label{fig:paramagnetic}
\end{center}
\vskip-0.5cm
\end{figure}

Single file diffusion, with a non-Fickian scaling of the \msd as $t^{1/2}$ is observed in both experiments and numerical simulations, allowing the determination of the mobility. A dependence of the reduced mobility as a function of the reduced interaction $\Gamma_{\rm para}$ is observed for both data sets (see Fig.~4 of \cite{Wei00} and Fig.~3-b of \cite{Herrera08}), summarized here in our Fig.~\ref{fig:paramagnetic}. 

The numerical simulations fulfill all assumptions of the theoretical calculations (see Table~\ref{tab:listhypo}), and their results are indeed in excellent agreement with the theory.  The small discrepancy between experiments and theory may be due to the  hydrodynamic interactions between the particles in the file. The experiments are done at constant density and constant temperature, hence the fact that the agreement is less good at small magnetic potential ($\Gamma_{\rm para} < 4$, typically) may be explained by the greater relative importance of hydrodynamic interactions compared to those between magnetic dipoles. Hydrodynamic interactions are basically repulsive, so that neglecting them lead to overestimating the compressibility of the brownian particles. This is consistent with the fact that the SFD mobility measured in \cite{Wei00} is less than what is expectated without hydrodynamic interactions. In the experiments of Lutz \emph{et al.} \cite{LutzPRL04}, great care have been taken to confine the beads in optically designed channels, in order to minimize hydrodynamic interactions. The quantitative agreement with Kollmann's results seems better in this configuration.

\subsubsection{Screened electrostatic interactions}
\label{sec:elec}

In another set of numerical simulations \cite{Herrera07}, Herrera-Velarde and Casta$\tilde{\rm n}$eda-Priego study colloidal particles with screened electrostatic interactions, described by the DLVO pair potential \refpar[eq:gammaDLVO]. They vary the packing fraction $\varphi$ of the brownian particles, which is linked to the density $\rho$ by $\rho = \varphi/\sigma$, where $\sigma$ is the particles diameter. They observe SFD behavior, with a non-fickian scaling for the \msd (see their Fig.~2-a), and measure the SFD reduced mobility \refpar[eq:mobreduced] as a function of the colloid packing fraction $\varphi$. 

\begin{figure}[htb]
\begin{center}
\includegraphics[scale=1]{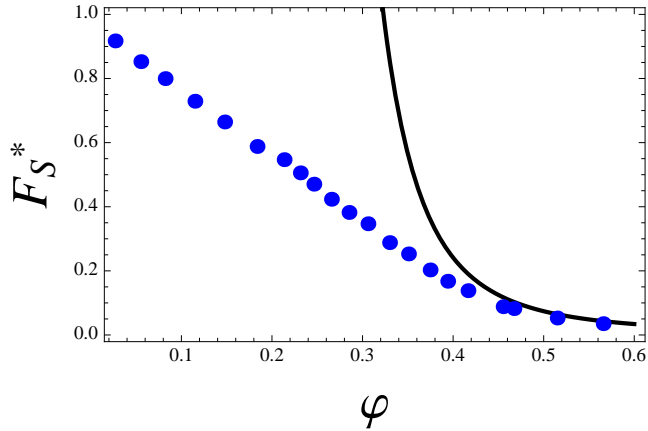}
\caption{(color online) Plot of the reduced mobility $F_S^*$ as a function of the brownian particles packing fraction $\varphi$, for colloids with screened electrostatic interactions (DLVO). \textcolor{blue}{$\bullet$}~: Numerical simulations \cite{Herrera07}, solid line Kollmann's formula, Eqn.~\refpar[eq:mobreduced] with \refpar[eq:gammaDLVO] (no free parameter in this calculation).}
\label{fig:electric}
\end{center}
\end{figure}

In Fig.~\ref{fig:electric}, we compare the simulations data  (see Fig.~2-b of \cite{Herrera08}) to the theoretical prediction of Kollmann. The agreement is very good at high packing fraction ($\varphi > 0.4$), and very poor at lower ones. A possible issue explaining the discrepancy at low packing fraction, is that the time taken to access a fully developed non-fickian scaling for the \msd may be large compared to the duration of a simulation. When particles are organized sequentially on a line, their diffusion is normal (Fickian) at small time, and they must cover diffusively a lengthscale comparable to their mean distance in order to feel the presence of other beads and undergo SFD. SFD cannot be observed before a  typical time of order $a^2/D_{eff}$ where $a$ is the interparticle mean distance and $D_{eff}$ the diffusivity of interacting particles (see below section~\ref{sec:crossover}). In simulations as in experiments, it means that the measurement duration has to be greater at smaller packing fraction. Indeed, as seen from Fig.~2-a of \cite{Herrera07}, at a packing fraction of 0.288 the $t^{1/2}$ scaling is not very good, and the reduced mobility measurement may not be very good either. This effect should be worse at lower packing fraction. 

Involving the smallness of the packing fraction is consistent with other results. The results reported in Fig.~\ref{fig:paramagnetic} are obtained with paramagnetic colloidal particles, at a packing fractions of 0.33. For our own system, the packing fraction is always greater than 0.38. As shown by Fig.~\ref{fig:mobility} or Fig.~\ref{fig:mobiltemp}, the agreement with Kollmann's result is much better than for simulations of low packing fraction colloids (Fig.~\ref{fig:electric}). 

\subsubsection{Discussion}
\label{sec:concpar}

We have done in this section a quantitative comparison between experimental and numerical studies of systems involving SFD, and a theory that takes into account finite range interactions between the diffusing particles \cite{Kollmann03}. The qualitative behaviors, predicted by the theoretical analysis, are indeed observed~: At large time, non-fickian diffusion occurs with the peculiar SFD subdiffusive behavior $\langle\Delta x^2 \rangle= F  {t}^{1/2}$ for the mean squared position fluctuations. The mobility $F$ decreases when the brownian particles compressibility decreases, either because of an increasing interaction strength (\S~\ref{sec:para}) or of an increasing packing fraction (\S~\ref{sec:elec}). This shows that the assumptions at the basis of the theoretical approach are rather consistent with actual realizations of SFD (see Table~\ref{tab:listhypo}). 

Quantitatively, the test is stringent because once the potential is known, then the reduced mobility is determined without any free parameter.  For colloidal particles interacting as magnetic dipoles \cite{Wei00,Herrera08}, at a packing fraction of 0.33, the agreement of theory with simulations is excellent, and very good with the experiments (see Fig.~\ref{fig:paramagnetic}). For screened electrostatic (DLVO) interactions, the simulations are done with varying packing fraction  \cite{Herrera07}, and the agreement is good at high packing fraction only (see Fig.~\ref{fig:electric}).

In the next sections, we do this quantitative comparison for a versatile system (see \ref{sec:setup}) of macroscopic charged  beads in which the interaction strength, temperature, density  and system size may be independently varied. The interaction potential is known from previous studies \cite{Galatola06}, and the comparison again involves no free parameter. The time resolution and experiment duration is sufficient to describe clearly both the short time regime, intermediate between ballistic transport and normal (Fickian) diffusion, and the long time regime for which we exhibit SFD behavior. We discuss the crossover time between those regimes, and show that, in order of magnitude, it is the time taken by a tagged particle to explore diffusively the potential well created by its two neighbors. When the SFD is observed, we measure the mobility and discuss its variations with the potential strength, the density and the temperature. We pay particular attention to potential finite size effects and show that, even for quite small systems, they do not intervene because of the periodic boundary conditions. This is a solid evidence that calculations done in the thermodynamic limit, hence infinite system size, are actually relevant.

\section{A system of macroscopically charged particles}
\label{sec:process}
\subsection{Experimental set-up}
\label{sec:setup}

Our experimental set-up consists in millimetric stainless-steel beads (diameter 0.8~mm, mass $m=2.15$~mg) located on a silicon wafer which is the bottom electrode of a plane horizontal condenser. A metallic frame intercalated between the electrodes, in contact with the bottom one, confines the beads in a circular channel of radius $R$  in such a way that they cannot cross each others. The bead motions may be decomposed in radial and orthoradial coordinates, wich have been shown to be mutually statistically independent \cite{Coupier06}. Since the annulus radius of curvature is much greater than the beads mean distance, we have single file diffusion with periodic boundary conditions. We have done experiments with two annulus, one radius $R =  4$~mm and width 2~mm, the second of radius $R =  9$~mm and width 2~mm. We report on experiments done with $N = 12,\; 14$ or $16$ beads in the small annulus (densities $\rho = 477$, 557 and 637 particles per meter), and with $N = 27,\; 32$ or $37$ beads in the large one ($\rho = 477$, 566 and 654 particles per meter).   

When a voltage $V_0 \in [0.5,1.3]$~kV is applied between the two electrodes, the beads become charged and repel each other with a known interaction \cite{Galatola06}  [see above eqn. \refpar[eq:gammabeads]]. In order to introduce an effective thermodynamic temperature, the condenser is fixed on loudspeakers excited with a white noise voltage. We have verified that the mechanical shaking, transmitted to the beads via the friction with the bottom electrode, does indeed behave as a thermal noise. This point is discussed in the following section~\ref{sec:temp} and in a preceding work \cite{Coupier05}.

\begin{figure}[htb]
\begin{center}
\includegraphics[scale=0.3]{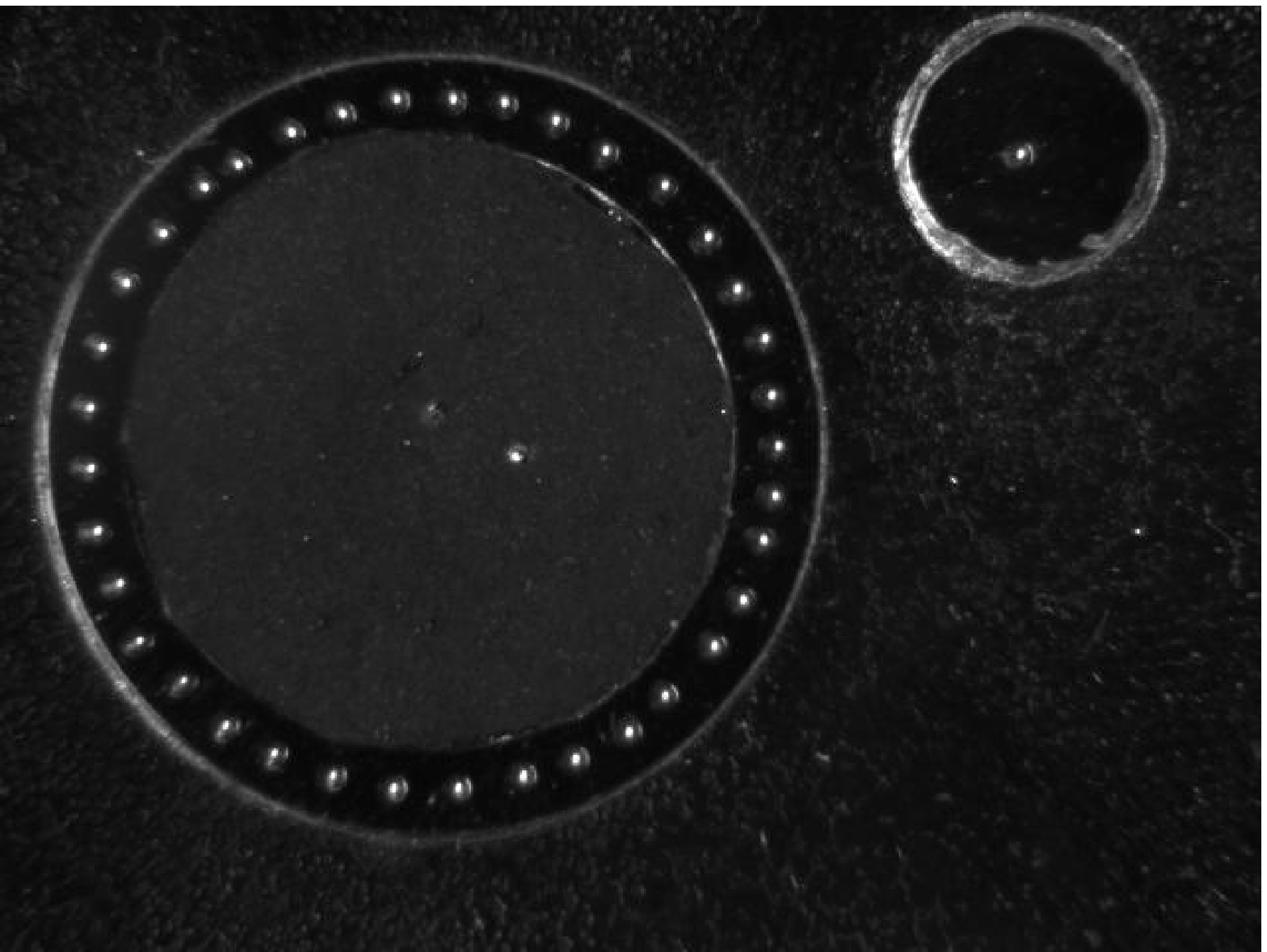}
\caption{\small Picture of the circular channel with $N = 37$ beads inside, recorded during an experimental run ($T = 10.2\;10^{11}$~K, $V_0 = 1000$~V). Inner radius 8~mm, outer radius 10~mm, beads diameter 0.8~mm. In the right-up corner, an identical bead confined in a disk of radius 5~mm is the embedded thermometer.}
\label{fig:photo}
\end{center}
\vskip-0.5cm
\end{figure}

To avoid any gravity effects on the beads, we have to ensure precisely the horizontality of the annulus. The experimental cell is supported by three columns, the extremities of which define a plane, the vibrations that mimic thermal excitation are exerted in that same plane, and we ensure a priori the horizontality of this plane. However, no inclination measurement is more sensitive than the system motion itself~: the mean particle distance should be uniform along the annulus for a satisfactory horizontality, which is checked before each experimental run. The picture displayed in Fig.~\ref{fig:photo} shows such a situation. 

The top electrode is a glass plate covered with an ITO (Indium-Tin-Oxyde) layer which is sufficiently fine (thickness 0.1~$\mu$m) to ensure optical transparency. Images of the particles may thus be recorded in real time during the experiments. A typical snapshot with 37 beads is provided in Fig.~\ref{fig:photo}. The interval between two snapshots ranges between 10 and 30~ms and series of 10 000 pictures were recorded. The effective thermal bath is characterized by a relaxation time about 100~ms \cite{Coupier06}, hence individual trajectories of the beads are determined over long times. The diffusion of the beads in an annulus of radius $R$ is measured through the time evolution of their orthoradial \msd $R^2\langle\Delta \theta^2(t)\rangle$ given by
\begin{equation}
R^2\langle\Delta \theta^2(t)\rangle = R^2\langle \biggl[\theta(t + t_0) - \theta(t_0) - \langle \theta(t + t_0) - \theta(t_0) \rangle \biggr]^2\rangle
\label{eq:defmsd}
\end{equation}
where $\theta$ is the orthoradial cumulated angle in radians and $t_0$ an arbitrary initial time. The bracket $\langle \cdot \rangle$ denote ensemble averaging. This averaging is done on every beads, since they all play an equivalent role. Moreover, the phenomenon is assumed to be stationary, so that $\langle\Delta \theta^2(t)\rangle$ do not depend on $t_0$. For a given time $t$, it makes thus sense to average on the initial time $t_0$. Let $n$ be the overall number of data recorded in one experimental run,  $\delta t$ the sampling time, and $n_t = t/\delta t$. Then the averaging on the initial time $t_0$ reads
\beq
\langle\langle\Delta \theta^2(t)\rangle\rangle = \sum\limits_{i = 0}^{n - n_t}{\left\{\theta[(n_t+i)\delta t] - \theta(i \delta t)\right\}^2 \over n - n_t + 1} -\left( \sum\limits_{i = 0}^{n - n_t}{\theta[(n_t+i)\delta t] - \theta(i \delta t) \over n - n_t + 1}\right)^2,
\label{eq:calcavg}
\eeq
where the index $i$ is such that $t_0 = i \delta t$. This way of averaging greatly improves the statistics when $n_t$ is smaller than $n$. In what follows, this double averaging will be denoted $\langle\Delta \theta^2(t)\rangle$ for simplicity.

\subsection{The effective temperature}
\label{sec:temp}

\begin{figure}[htb]
\begin{center}
\includegraphics[scale=0.8]{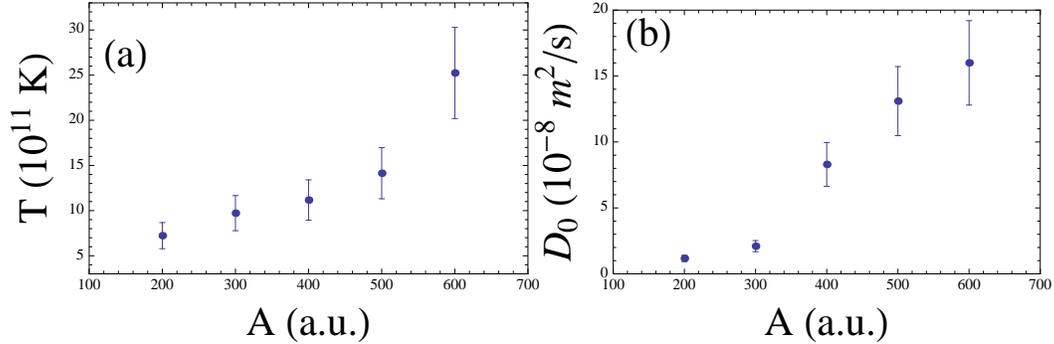}
\caption{(a) Effective thermodynamic temperature $T$ (in $10^{11}$~K) as a function of mechanical agitation $A$ (a.u.). (b) Single particle diffusivity $D_0$ ($10^{-8}$~m$^2$~s$^{-1}$) for orthoradial motions in an annulus, as a function of mechanical shaking amplitude $A$ (a.u.). In each graph, error bars are determined statistically from several measurements, and represent typically $\pm 5 \%$ of the mean value.}
\label{fig:tempdiff}
\end{center}
\vskip-0.5cm
\end{figure}

It has been carefully checked in previous works \cite{Coupier05, Coupier06} that the mechanical agitation is, indeed, equivalent to a thermodynamical temperature. In \cite{Coupier05}, the random position of a single bead rolling on a wafer inclined by a very small angle was recorded, and its probability distribution compared to the Boltzmann distribution for several shaking amplitudes $A$. An excellent agreement was obtained, allowing the definition of an effective temperature $T$ which depends on the shaking amplitude. Then, an \emph{in situ} thermometer was built, made from a single bead trapped in a circular frame located near the main one and submitted to the same voltage. Such a thermometer is shown in Fig.~\ref{fig:photo}. The bead is confined in a 2D parabolic potential. As it should, its radial mean square displacement is a linear function of the effective temperature $T$ in the long time limit, hence this set-up actually behaves as a thermometer. The temperature calibration is shown in Fig.~\ref{fig:tempdiff}~(a). The voltage $V_0 = 1000$~V was kept constant during all measurements. The error bars (roughly $\pm 5\%$ of the mean value) reflect the statistical dispersion of the measured value for several experiments undertaken at the same shaking amplitude $A$. 

The mobility of a tagged particle in a file depends on the single particle diffusivity $D_0$, both for hard spheres  \refpar[eq:mobilhard] or finite range interactions. In this latter case,  the effective diffusivity $D_{eff}$ is a function of $D_0$, the temperature and the compressibility \refpar[eq:mobilsoft]. In \cite{Coupier06}, the diffusion of a single bead in a circular channel was studied.  It was shown that the radial and orthoradial motions are completely decoupled. The orthoradial motion is well described by a free diffusion with diffusivity $D_0$, whereas the radial motion is that of a brownian particle in a 1D parabolic potential. The diffusivity $D_0$ has been measured as a function of the mechanical shaking amplitude $A$, at constant applied voltage. As shown by  Fig.~\ref{fig:tempdiff}~(b), it is a monotonously increasing function of $A$.

The mechanical shaking exerted on the beads acts as an \emph{effective thermal bath}, and it is thus possible to describe the beads motion with the help of a Langevin equation \cite{Coupier06}. The complicated coupling, due to solid friction, between the horizontal vibrations of the silicon wafer and the bead motions behaves as the solvent particules that exert brownian random forces on a colloid. However, the shaking is physically different from the erractic motion of solvent molecules around colloidal particles, and hydrodynamic interactions are avoided in our system. 

\section{Single file diffusion of macroscopic charged particles}
\label{sec:results}
\subsection{Single file diffusion}
\label{sec:sfd}

\begin{figure}[htb]
\begin{center}
\includegraphics[scale=1]{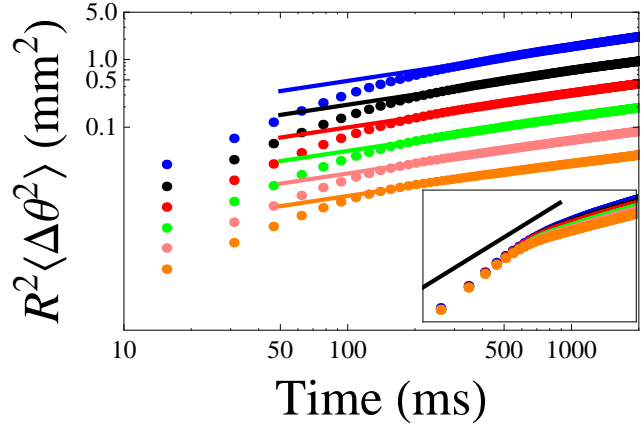}
\caption{(Color online) Plot of the orthoradial m.s.d. $R^2\langle\Delta \theta^2\rangle$ (in mm$^2$) as a function of time $t$ (in ms), in a log-log scale, for $\rho = 654$~m$^{-1}$ (37 beads, $R = 9$~mm), and $T = 10^{12}$~K. The applied voltage $V_0$ vary from $900$~V to 1400~V in steps of 100~V in order blue-black-red-green-pink-orange. For clarity, each data set have been shifted by Log(2) except for $V_0 = 1400$~V. The solid lines are the best fits with \refpar[eq:SFD], where the mobility $F$ is the adjustable parameter. The inset display the unshifted data, showing that the diffusion at small time is independent on the voltage. The solid line is of slope 1. The dimensionless potential energy is such that $16 < \Gamma_{\rm beads} < 40$.}
\label{fig:scaling}
\end{center}
\vskip-0.5cm
\end{figure}

 In Fig.~\ref{fig:scaling}, we plot the  orthoradial m.s.d. $R^2\langle \Delta \theta^2 \rangle$ as a function of time, for $\rho = 654$~m$^{-1}$, $T = 10^{12}$~K, and varying applied voltage $V_0$, hence varying interaction strength [see \refpar[eq:gammabeads]]. The behavior of a bead in a single file geometry depends on the observation time. At very small time, the bead motion is ballistic, with a $\hbox{\msd} \propto t^2$. At intermediate times, normal (Fickian) diffusion with $\hbox{\msd}\propto t$ takes place, because some time is needed for the statistics of the particle to be influenced by the single file geometry. As shown by the inset, in this regime the \msd is somewhat faster than linear, intermediate between ballistic transport and ordinary diffusion. This behavior is independent of the applied voltage.   
 
For brownian particles confined on a line,  SFD may occur with non-fickian scaling $\hbox{\msd}\propto t^{1/2}$ for asymptotically long times. This is what we observe at times greater than typically 200~ms, as shown by Fig.~\ref{fig:scaling}, for all voltage values. The inset gives evidences that the mobility actually depends on the applied voltage, hence on the interparticle interactions. This dependence is studied in detail in section \ref{sec:mobil}. The main plot shows that the quality of the fit is good. It evidences a crossover between a normal diffusion regime, and the appearance of SFD behavior. We may thus define a crossover time $\tau_c$, which is shown to increase with decreasing applied voltage, hence decreasing interparticles interaction [see \refpar[eq:gammabeads]] or increasing compressibility [see \refpar[eq:compressibility]]. The same behavior was already observed in colloids (see \cite{Wei00}, Fig.~2-B). A more quantitative analysis of the crossover time is the subject of section~\ref{sec:crossover} below.

\subsection{Mobility measurements}
\label{sec:mobil}

\begin{figure}[htb]
\begin{center}
\includegraphics[scale=0.75]{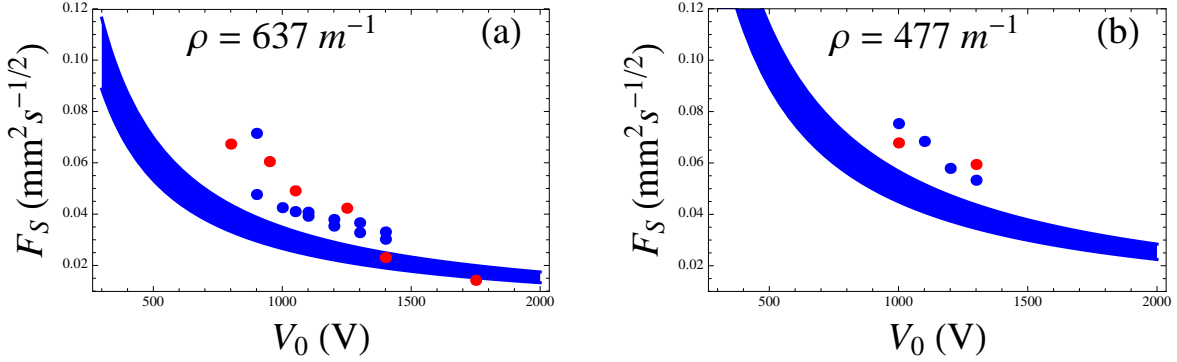}
\caption{(a) Mobility $F_S$, in mm$^2$s$^{-1/2}$ as a function of the applied potential $V_0$ in V. Blue dots, $N = 37$, $R = 9$~mm; red dots, $N = 16$, $R = 4$~mm. The solid lines are the theoretical expectations for temperatures $T = 10.6\;10^{11}$~K and $T = 11.7\;10^{11}$~K (no free parameter). Dimensionless potential energy~: $11 < \Gamma_{\rm beads} < 55$. (b) Blue dots, $N = 27$, $R = 9$~mm; red dots, $N = 12$, $R = 4$~mm. The solid lines are the theoretical expectations for temperatures $T = 10.7\;10^{11}$~K and $T = 10.2\;10^{11}$~K (no free parameter). Dimensionless potential energy~: $6 < \Gamma_{\rm beads} < 10$.}
\label{fig:mobility}
\end{center}
\end{figure}

The $t^{1/2}$ scaling of the \msd, typical of SFD, is clearly observed in Fig.\ref{fig:scaling}, allowing mobility measurements using \refpar[eq:SFD] with the mobility as the fitting parameter. We vary the applied voltage and the mechanical agitation, and to increase the statistics, we record several picture sets at the same voltage and shaking amplitude. We display in Fig.~\ref{fig:mobility} the mobility as a function of the applied voltage, for $\rho = 637$~m$^{-1}$ [Fig.~\ref{fig:mobility}-(a)] and $\rho = 477$~m$^{-1}$ [Fig.~\ref{fig:mobility}-(b)], at fixed shaking amplitude. As shown by the error bars in Fig.~\ref{fig:tempdiff}, there is some uncertainty on the temperature for a given mechanical agitation level. It means that, when we want to compare our mobility measurements to the theoretical prediction \refpar[eq:mobilsoft], the error done when evaluating the temperature $T$ and diffusivity $D_0$, knowing the shaking amplitude $A$, prevails on any other one. We thus plot the prediction from \refpar[eq:mobilsoft] for the lowest and highest temperatures consistent with amplitude $A$. 

The data clearly exhibit qualitative features that are in excellent agreement with the theoretical predictions, at both densities. The mobility depends on the strength of the interparticle interaction, showing that SFD cannot be described by the mobility \refpar[eq:mobilhard], valid for hard sphere interactions. The mobility decreases when the voltage increases. This is fully consistent with \refpar[eq:mobilsoft], because increasing $V_0$ means increasing the interaction energy and decreasing the compressibility, as shown by \refpar[eq:gammabeads] and \refpar[eq:compressibility], hence decreasing the mobility as stated by \refpar[eq:mobilsoft]. Quantitatively, the agreement is correct. An important point, that will be developed in the next section, is that the system is described by its density rather than by its size, because the mobility values taken with small and large annulus are consistent at a given density.

\begin{figure}[htb]
\begin{center}
\includegraphics[scale=1.]{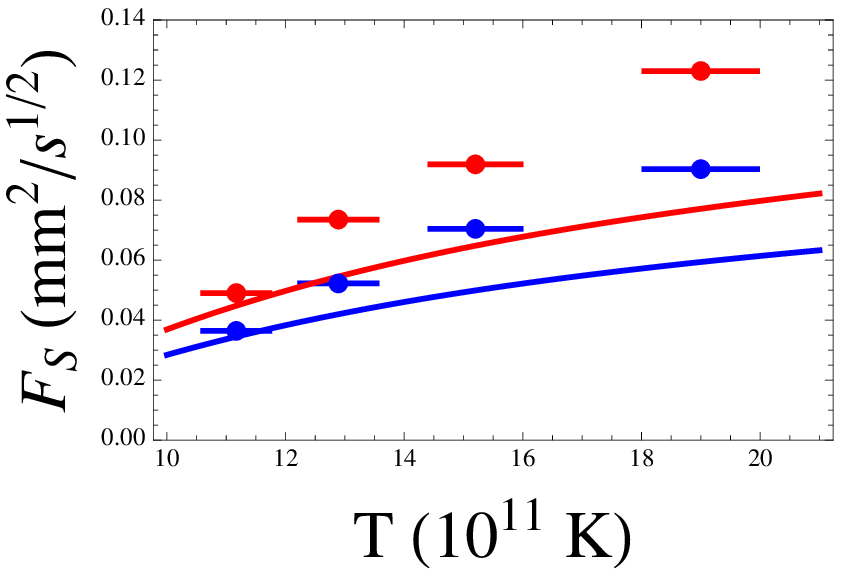}
\caption{ Mobility $F_S$, in mm$^2$s$^{-1/2}$ as a function of the temperature ($\pm 5\%$ error bar), for a density of ${\rho} \approx 561$m$^{-1}$. Blue error bars : $V_0 = 1300$~V and $11 < \Gamma_{\rm beads} < 19$, red error bars : $V_0 = 1000$~V and $6 < \Gamma_{\rm beads} < 11$. The solid lines follow the same color code, and are the theoretical expectations for the mean temperature.}
\label{fig:mobiltemp}
\end{center}
\end{figure}

In Fig.~\ref{fig:mobiltemp}, we display the mobility as a function of temperature, for a constant density of 561~m$^{-1}$, and two voltages. Since the diffusivity $D_0(T)$ is an increasing function of the temperature, the mobility should increase with the temperature, and it should be higher at lower applied voltage since the compressibility decreases with the voltage. At a given temperature, the mobility measured for $V_0 = 1300$~V is indeed always smaller than for $V_0 = 1000$~V, in good qualitative agreement with the theoretical picture. However, the quantitative agreement is not very good.

In order to interpret those discrepancies, two possible issues may be considered. The first one is the smallness of the beads number, whereas calculations are done in the thermodynamic limit. The larger beads numbers in our experiments (27, 32 and 37) are nevertheless quite similar with those in previous experiments (see Table~\ref{tab:listhypo}). 

Another issue is suggested by Nelissen \emph{et al.} \cite{Nelissen07} to interpret the discrepancies with SFD behavior that they observe in their simulations. Their algorithm use the Langevin equation with DLVO interaction between particles without assuming overdamped dynamics, taking into account the inertial effects (acceleration). Overdamped dynamics is at the basis of both Kollmann's calculation, and the algorithm used in  \cite{Herrera07,Herrera08}.  It is a safe assumption in colloids \cite{Nagele96}, but  our system is clearly not in this regime, as shown in \cite{Coupier06}. When the radial \msd of a bead in a parabolic potential is compared to the predictions of a Langevin equation (see the Fig.~4 and Fig.~5 of \cite{Coupier06}), it is clear that the bead undergoes oscillations, that are \emph{not} overdamped since a pseudoperiod should easily been defined. We may get an intuitive idea of the corrections due to inertial effects  as follows. If the particle motion is not overdamped, it declines more slowly than in an overdamped regime. It may explain why we always observe a greater mobility than the prediction obtained assuming overdamping. Another satisfactory issue is that the agreement is better either at low temperature (the lowest temperature in Fig.~\ref{fig:mobiltemp} is in very good agreement with the theory) or at large applied voltage (hence large interaction), when thermal motion is comparatively less important. 

\subsection{Finite size effects}
\label{sec:size}

The theorerical description of SFD \cite{Kollmann03} takes place in the thermodynamic limit, for infinite particle number $N$, system size $L$, and finite mean density $\rho \equiv N/L$. In either experiments or simulations, the particle number is obviously finite, and boundary effects are avoided using periodic boundary conditions, putting the diffusing particles in an annulus. However, a finite size effect that cannot be avoided is the fact that an acoustic mode propagating along a ring, comes back to its point of departure after one turn rather than going away at infinity. Such modes, obviously missing in infinite systems,  may have an effect on the correlations that are responsible for the SFD.  It is thus of interest to look at the behavior of systems with the same density, but different sizes. 

\begin{figure}[htb]
\begin{center}
\includegraphics[scale=0.8]{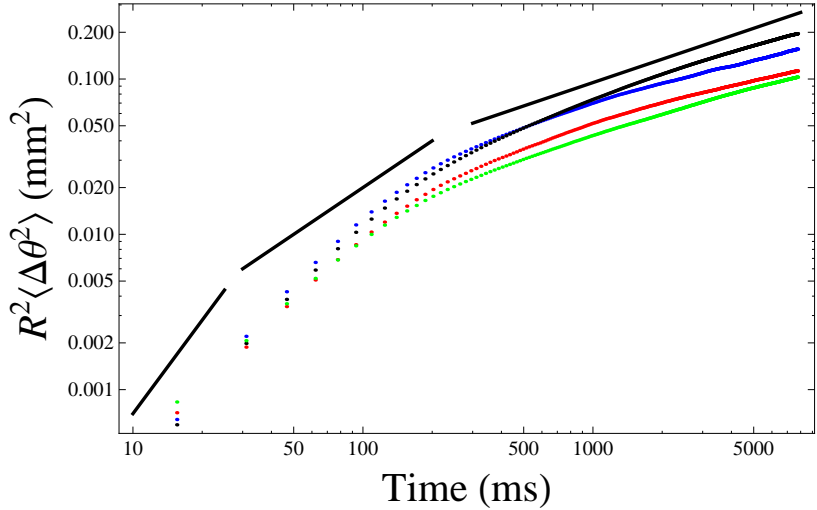}
\caption{(Color online) Plot of the orthoradial \msd $R^2\langle\Delta \theta^2\rangle$ (in mm$^2$) as a function of time $t$ (in ms), in a log-log scale, for densities 477~m$^{-1}$ (blue dots, small annulus, $N = 12$; black dots, large annulus, $N = 27$) and 637~m$^{-1}$ (red dots, small annulus, $N = 16$; green dots, large annulus, $N = 37$). The solid black lines are of slopes $2$, $1$ and $1/2$. The temperature is the same in all experiments ($T = 11.2\;10^{11}$~K) and the applied voltage is 1000~V. The crossover time decreases when the density increases.}
\label{fig:crossover}
\end{center}
\end{figure}

In Fig.~\ref{fig:mobility}, we display mobility measurements  done at two densities, both with large and small annulus. The smaller system contains few beads (12 and 16), but the data are nevertheless in good agreement with those for larger bead numbers and the same density. The density is thus the relevant parameter, and finite size effects are not exhibited by the measurements. In order to test potential finite size effects on more sensitive data, we display in Fig.~\ref{fig:crossover} the rough recording of orthoradial \msd as a function of time. All data are obtained at the same temperature, as measured by the embedded thermometer, same applied voltage and two different densities~: 477 beads per meter with $N = 12$ in the small annulus and $N = 27$ in the large one, 637 beads per meter with $N = 16$ in the small annulus and $N = 37$ in the large one. 

The time evolution of the \msd is the same for a given density, whatever the particles number. At large time, the \msd evidences SFD behavior in each case. As may be expected, the systems with smaller density exhibit the larger \msd. All four data sets are necessarily recorded separately, because they involve a complete dismantling and reassembly of the experimental cell. They thus show a very satisfactory reproducibility, and provide a convincing evidence that the system is correctly described by its density, without finite size effects. 

As was already the case for the data reported in fig.\ref{fig:scaling}, the small time evolution of the \msd is faster than linear, between ballistic transport and ordinary diffusion. At larger times, non-fickian SFD with $t^{1/2}$ behavior is recovered. It is clear that the crossover time between normal diffusion and SFD increases with decreasing density. This is satisfactory, because a smaller density means a larger distance between the particles, hence a longer time to feel their neighbors and undergo SFD.

\subsection{Crossover time}
\label{sec:crossover}

In this section, we focus on the crossover time $\tau_c$ at which the system evolves from ordinary diffusion toward SFD. Since this evolution is continuous, the crossover time is roughly defined, in terms of order of magnitude only. This crossover time  is the time needed by a given particle to explore a lenghscale sufficiently large to "feel" that it is in a sequence, and that its diffusion is perturbed by the neighboring particles. The relevant diffusivity is not $D_0$, that of a single particle in a thermal bath, but rather the renormalized diffusivity $D_{eff}$ which takes into account the interactions between brownian particles \cite{Nagele96}. Neglecting hydrodynamic interactions between particles, we get $D_{eff} = D_0/S(0,0) = D_0/(\rho k_B T \kappa_T)$. We thus expect the crossover time to be such that
\beq
\tau_c \rho^2 D_{eff} = \tau_c \rho^2 {D_0\over \rho k_B T \kappa_T}\sim 1.
\label{eq:crossover1}
\eeq
Using the relationship between the interaction potential and the compressibility \refpar[eq:compressibility], we may rewrite this expression as 
\beq
\tau_c  D_0{U'' \over  k_B T }\sim 1.
\label{eq:crossover2}
\eeq
We can thus give another interpretation of the occurence of the effective diffusivity in \refpar[eq:crossover1]. The motion of a tagged particle may be described as a diffusion with diffusivity $D_0$, in the potential  $U(r)$ due to the interactions exerted by its two neighbors (neglecting the obvious fact that the walls of this potential well are themselves fluctuating). If the particle is in a thermal bath at temperature $T$, at large times its typical \msd in this potential well is $k_B T/U''$. The crossover time $\tau_c$ is such that the particle "knows" that it is in a sequence, hence that it has felt the walls of the potential well due to its neighbors. This is exactly what is expressed by \refpar[eq:crossover2].

The compressibility may be calculated from \refpar[eq:compressibility] and the relevant expressions for the interaction potentials \refpar[eq:gammaDLVO], \refpar[eq:gammapara] and \refpar[eq:gammabeads]. Several papers provide evidences of a transition between normal diffusion and SFD, in such a way that a crossover time may be estimated in each case \cite{Wei00,LutzPRL04,Lin05,Herrera07, Herrera08}. Moreover, we deduce the crossover time for macroscopically charged beads from Fig.~\ref{fig:scaling}. Those estimates are summarized in Table~\ref{tab:crossover}. We point out that the dimensionless quantity $\tau_c \rho^2 D_{eff}$ is in each case roughly of order one, although in their respective physical units, the numerical values for all quantities $\tau_c$, $\rho$ and  $D_{eff}$ differ by several orders of magnitudes in colloids and in our experiments.

\begin{table}[htb]
\begin{center}
\begin{tabular}{|c|c|c|c|c|c|}
\hline
Reference & $\quad \tau_c \quad$ & $\rho^{-1}$ & $D_0$ & $\quad \tau_c \rho^2 D_0 \quad$ &  $\quad \tau_c \rho^2 D_{eff} \quad$\\
\hline
\hline
\cite{Wei00}, Fig.~2(B) & 10~s & 11~$\mu$m & 0.036~$\mu$m$^2$s$^{-1}$ & $3\; 10^{-3}$ & 0.32 \\
\hline
\cite{Herrera07}, Fig.~2(a) & -- & 5.7~$\mu$m & -- & $2\; 10^{-2}$ &1.16 \\
\hline
\cite{Herrera08}, Fig.~3(a) & -- & 6.1~$\mu$m & -- & $4\; 10^{-3}$ &0.39 \\
\hline
Fig.~\ref{fig:scaling} & 0.2~s & 1.6~mm & 0.02~mm$^2$s$^{-1}$ & $1.6\; 10^{-3}$ &0.44 \\
\hline
\end{tabular}
\caption{Crossover time for several SFD observations. For colloids, the observed crossover times in \cite{LutzPRL04} (Fig.~2) and \cite{Lin05} (Fig.~2) are quite comparable to those of \cite{Wei00}, for roughly the same densities, temperature and particle sizes.}
\label{tab:crossover}
\end{center}
\end{table}

\begin{figure}[htb]
\begin{center}
\includegraphics[scale=1]{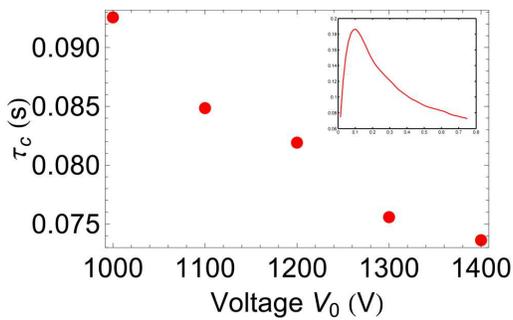}
\caption{(Color online) Crossover time $\tau_C$ (in s.) as a function of the applied voltage (in V.) In the inset we plot the \msd divised by the time as a function of the time. The crossover time is estimated as the position of the maximum of this curve.}
\label{fig:Ttransition}
\end{center}
\end{figure}

In Fig.~\ref{fig:Ttransition}, we display the evolution of the crossover time with the applied voltage. To this aim, we plot $R^2\langle\Delta \theta^2\rangle/t$ as a function of the time $t$. At very small time, the transport is ballistic, the \msd scales as $t^2$ and this function increases as $t$. At large time, SFD occurs with a \msd scaling as $t^{1/2}$, so that the function decreases as $t^{-1/2}$. When ordinary diffusion takes place, the function should exhibit a plateau. As shown by the inset of Fig.~\ref{fig:Ttransition}, this plateau is quite small. In fact, the function $R^2\langle\Delta \theta^2\rangle/t$ rather display a maximum, and the position of this maximum gives a rough estimate of the crossover time. The crossover time decreases with the applied voltage, which is in qualitative agreement with \refpar[eq:crossover2]. Quantitatively, the observed variation is weaker than what may be expected.

\section{Conclusion}
\label{sec:conclusion}

In this paper we discuss quantitatively experimental evidences of SFD in the transport of macroscopic charged beads, interacting with a screened electrostatic potential. To ensure periodic boundary conditions, the beads are aligned in an annulus narrow enough to avoid any crossing. For a time resolution of 10~ms and experimental runs duration about 1~h, the system timescales are such that we have access to both short time dynamics, intermediate between ballistic transport and usual diffusion, and long time dynamics that evidences single file diffusion. 

In this regime, the orthoradial mean squared displacement is such that $R^2\langle\Delta \theta^2 \rangle= F  {t}^{1/2}$. For finite range interactions, it has been shown recently by Kollmann that the mobility $F$ depends on the brownian particles compressibility \cite{Kollmann03}. When the interaction potential, hence the compressibility, is explicitely given, the mobility calculation does not involve any free parameter. All qualitative features displayed by our mobility measurements are in excellent agreement with the theory~: The SFD mobility decreases whith the interaction potential (Fig.~\ref{fig:mobility}), decreases with the brownian particles packing fraction, and increases with the temperature (Fig.~\ref{fig:mobiltemp}).  Quantitatively, we observe a SFD mobility that is somewhat greater than the theoretical predictions. We interpret this discrepancy using an argument taken in \cite{Nelissen07}. Those authors point out that a key assumption of the theoretical calculation is that the particle dynamics is  overdamped, which is not the case in our system. 

The theoretical calculations are done in the thermodynamic limit. All systems actually exhibiting SFD involve a finite number of particles, and periodic boundary conditions. In experiments, the particles number may be quite small. Using two different experimental cells, we show that the data recorded in small systems (12 to 16 particles) are consistent with those for larger systems (27 to 37 beads), and that experiments done at the same temperature are quite reproducible (Fig.\ref{fig:crossover}). Since all systems in which SFD may be a practical issue (\emph{e.g.,} ion transport in biological membranes \cite{Hodkin55,Rosenberg78,Hernandez92}) presumably involve few diffusing particles, this constitutes a satisfying evidence of the relevance of Kollmann's calculation. 

In our experiments, we clearly observe both the short time behavior of the system, where ballistic transport give way to ordinary (Fickian) diffusion, and the long time behavior where correlations led to SFD. We are thus able to define a crossover time $\tau_c$ between those two regimes. This may be done, perhaps less convincingly, in colloids where the short time regime is quite hard to exhibit.  The relevant physical parameters differ by several orders of magnitude in our systems and in colloids. Nevertheless, the dimensionless quantity  $\tau_c \rho^2 D_{eff}$ is of order one in each system, where the effective diffusivity $D_{eff}$ takes into account the interparticles interactions (see \S~\ref{sec:crossover}). This is in good agreement with the intuitive picture of SFD, which takes place when a tagged particle have had the time to explore diffusively its environment, and feel the presence of  the neighboring particles, placed at an average distance $1/\rho$.

We review previous numerical  \cite{Herrera07, Herrera08} and experimental  \cite{Wei00} data on colloids that evidence SFD behavior. We recast those data and provide a quantitative comparison with the theory for finite range interactions. All qualitative features predicted by the theory are observed (Fig.~\ref{fig:paramagnetic} and Fig.~\ref{fig:electric}). Quantitatively, the agreement is excellent for not too small packing fraction, as shown by Fig.~\ref{fig:paramagnetic}. The numerical algorithms fulfill all theoretical assumptions (see Table~\ref{tab:listhypo}), and hydrodynamic interactions may explain the small discrepancies between experiments \cite{Wei00} and theory. At smaller packing fraction, the numerical data do not agree with theory (Fig.~\ref{fig:electric}), which is presumably due to an insufficient simulation duration.

The calculations of Kollmann \cite{Kollmann03} rely on two fundamental assumptions~: overdamped dynamics and no hydrodynamic interactions. Those latter are either avoided or negligible in the works reported here. However, in some practical situations (transport through molecular-sized channels in biological membranes, molecular sieving effects in nanoporous zeolites, diffusion of colloids in microfluidic devices), hydrodynamic interactions may play a significant part in the dynamics.  In the diffusion of molecules along folded polymer chains, it is rather the overdamping assumption that may be incorrect. We have shown that it should give observable corrections of the SFD mobility. Theoretical studies should thus be pursued in those directions, but as it states, the theory of Kollmann that takes into account the softness on the interparticle interactions constitutes a very significant improvement on previous approaches assuming hard sphere interactions only.

\leftline{\bf Acknowledgments}

We thank C. Guthmann for many discussions, and careful reading of the manuscript.

\bibliography{Biblio_D1D}
 \bibliographystyle{unsrt}

\end{document}